\documentclass[12pt]{article}
\usepackage{mathrsfs}
\usepackage{amsmath,amssymb,array}
\usepackage{graphicx}

\numberwithin{equation}{section}
\def\p{\partial}

\begin{document}

\begin{titlepage}
\renewcommand{\thefootnote}{\fnsymbol{footnote}}

\begin{center}

\begin{flushright} \end{flushright}
\vspace{2.5cm}

\textbf{\Large{Mesons and Nucleons in Soft-Wall AdS/QCD}}\vspace{2cm}

\textbf{Peng Zhang} \\[0.5cm]

\textsf{E-mail: pzhang@bjut.edu.cn}\\[0.5cm]

\emph{Institute of Theoretical Physics, College of Applied Sciences, \\
      Beijing University of Technology, Beijing 100124, P.R.China}

\end{center}\vspace{1.5cm}

\centerline{\textbf{Abstract}}\vspace{0.5cm}
We study further the soft-wall AdS/QCD model with a cubic potential for the bulk scalar.
We analyze the spectra of pseudoscalar, scalar, vector and axial-vector mesons. We also
study the spin-1/2 nucleon spectrum and the pion-nucleon coupling. All of them have a
good agreement with the experimental data.

\end{titlepage}
\setcounter{footnote}{0}

\section{Introduction}

The low energy behavior of quantum chromodynamics (QCD) is very
important for people to understand various properties of the strong
interaction. This is a non-perturbative problem and there is no
systematical tool from the first principle to apply. Lattice
calculations can give us useful numerical results, but we still need
to understand many important phenomena, e.g. chiral symmetry
breaking and quark confinement, from analytical aspects.

Large $N$ expansion \cite{tH} is an excellent idea to deal with this
kind of strong coupling problems. By very general arguments, 't
Hooft showed that a gauge theory at large $N$ looks like some weak
coupling string theory. This idea was explicitly realized by the
Anti-de Sitter/Conformal Field Theory (AdS/CFT) duality \cite{M,
GKP, W}, which says that $\mathcal{N}=4$ four-dimensional
supersymmetric Yang-Mills theory is equivalent to type IIB string
theory in the $AdS_5 \times S^5$ spacetime with five-form fluxes.
Now this kind of gauge/gravity correspondences have been applied to
many fields in theoretical physics.

After introduced in \cite{EKSS, DP1}, the AdS/QCD model gives us a
new methodology to study low energy properties of the strong
interaction. It is based on the phenomenological assumption that at
low energy QCD can be described by a five-dimensional theory living
in a slice of $AdS_5$ or some deformation of it. Surprisingly, this simple
model indeed capture many important characteristics of QCD, and can
be used to fit the experimental data quite well. Firstly people
studied the so-called hard-wall model in which the QCD scale is
introduced by an IR cutoff. It is relatively simple and main
features of chiral symmetry breaking can be realized. For instance,
scalar and pseudoscalar mesons were studied in \cite{DP2, GMTY}, and
tensor mesons are analyzed in \cite{KLS}. In addition to the meson
sector, baryons can also be realized in the hard-wall model. There
are mainly two ways to achieve this: one is by introducing a pair of
bulk fields with half-integer spin, see e.g. \cite{HIY, KLY, HKSY,
AHPS, Z1}; the other is through a skyrmion-type soliton \cite{PW1,
PW2}. To realize linear confinement for mesons the authors of
\cite{KKSS} introduced a soft-wall version of AdS/QCD by adding a
quadratic background dilaton instead of a sharp IR cutoff. By a WKB
argument they show that the meson spectrum grows as $m_n^2\sim n$,
consistent with the experimental data. In \cite{GKK}, by using a
quartic potential for bulk scalars, the authors succeeded in
disentangling the effects of explicit and dynamical breaking of
chiral symmetries in this soft-wall model. Many works have been
done. A partial list of them is \cite{CDJN}-\cite{VS}.

Although many properties of mesons have been produced in the
soft-wall model, nucleons are still studied within the hard-wall
model. In the framework of hard-wall, mesons and nucleons have
different IR cutoffs \cite{MT, KKY}. This problem becomes more
serious when considering their couplings. On the other hand,
\cite{Z1} has shown that higher nucleon states can be fit well by
using a relatively large IR cutoff. Therefore it seems very natural
to extend nucleons to a soft-wall model, which can properly realize
the asymptotically linear spectra for both mesons and
nucleons\footnote{There are already some works concerning this issue
in different models, see e.g. \cite{FBF, TB}.}, and consistently
encode the couplings between them. For these purposes, \cite{Z2}
suggested a cubic potential for the bulk scalar. Instead of the
dilaton for the meson sector, an asymptotically quadratic vacuum
expectation value (VEV) of the bulk scalar gives us the linear
spectrum for nucleons. The power of the potential then follows from
the relation between the dilaton and the expectation value through
the classical equation of motion. Since the bulk scalar is the
holographic dual of the quark bilinear operator, the cubic potential
will induce a three-point function of them at tree level and a
quantum correction of the quark condensate through a tadpole diagram
in the bulk.

In the present work, we will study this model further. By analyzing
the scalar meson sector, we reverse the sign in front of the the
coupling constant of the cubic potential. This allows us to fit the
data well not only for vector and axial-vector mesons as in our
previous paper, but also the scalar and pseudoscalar ones. We also
study the nucleon spectra and the effective pion-nucleon coupling
constant.

\section{Meson sector}

We first review some basic facts of the model. As in other soft-wall
AdS/QCD models, all fields are defined on a five-dimensional anti-de
Sitter space with the metric
\begin{eqnarray}
ds^2=\,G_{MN}\,dx^M dx^N=a^2(z)\,(\,\eta_{\mu\nu}\,dx^{\mu}dx^{\nu}-dz^2)\,,
\quad  0 < z < \infty\,,
\end{eqnarray}
where $a(z)=1/z$. The Minkowski metric
$(\,\eta_{\mu\nu})=\mathrm{diag}(+1,-1,-1,-1)$. There is also a
background dilaton $\Phi(z)$ which is assumed to be asymptotically
quadratic as $z\rightarrow\infty$. According to the general rules of
the gauge/gravity duality, there are two 5D gauge fields, $L_M^a$
and $R_M^a$, which are dual to 4D chiral currents
$J_{L}^{a\mu}=\bar{q}_L\gamma^{\mu}t^a q_L$ and
$J_{R}^{a\mu}=\bar{q}_R\gamma^{\mu}t^a q_R$. The quark bilinear
operator $\bar{q}_L^i q_R^j$ is also an important 4D operator for
chiral symmetry breaking. Its holographic dual is a 5D $2\times2$
matrix-valued complex scalar field $X=(X^{ij})$, which is in the
bifundamental representation of the 5D gauge group $SU(2)_L \times
SU(2)_R$. The bulk action for the meson sector is \cite{Z2}
\begin{eqnarray}
S_M=\int d^5x\, \sqrt{G}\,e^{-\Phi} \left\{-\frac{1}{4g_5^2}(\,\|F_L\|^2+\|F_R\|^2)+
    \|DX\|^2-m_X^2\|X\|^2+\frac{\lambda}{\sqrt{2}\,}\,\|X\|^3\right\}\,. \label{SM}
\end{eqnarray}
The norm of a matrix $A$ is defined as $\|A\|^2=\mathrm{Tr}(A^\dag
A)$. The cubic potential is crucial for realizing the asymptotically
linear spectra of both mesons and nucleons. Note that we have
reversed the sign in front of the coupling constant $\lambda$ of the cubic
potential as said in the introduction section. This will enable us
to fit the data of scalar mesons better. By matching with QCD's
results the parameters $g_5$ and $m_X$ are chosen to be
$g_5^2=12\pi^2/N_c=4\pi^2$ and $m_X=-3$. The covariant derivative of
$X$ is $D_M{X}=\p_M{X}-iL_M{X}+i{X}R_M$. The generator $t^a$ is
normalized by $\mathrm{Tr}\,t^at^b=\frac{1}{2}\delta^{ab}$. $F_L$ and
$F_R$ are the field strengths of the gauge potentials $L$ and $R$ respectively.
In the integration measure, $G$ denotes the determinant of the metric $(\,G_{MN})$,
so $\sqrt{G}=a^5$ for the $AdS_5$ space.

\subsection{Bulk scalar VEV and dilaton background}

The bulk scalar $X$ is assumed to have a $z$-dependent VEV as
follows
\begin{eqnarray}
\langle{X}\rangle=\,\frac{1}{2}\,v(z) \begin{pmatrix} 1 & 0 \\ 0 & 1 \end{pmatrix}.
\end{eqnarray}
The function $v(z)$ and the background dilaton $\Phi(z)$ are related through the following equation
derived from the bulk action (\ref{SM})
\begin{eqnarray}
\p_z(\,a^3 e^{-\Phi}\p_z v)-a^5 e^{-\Phi}(\,m_X^2 v-\frac{3}{4}\lambda\,v^2)=0\,. \label{EOMX}
\end{eqnarray}
As suggested in \cite{GKK}, we view it as an equation of the dilaton $\Phi$
\begin{eqnarray}
\Phi'(z)=\,\frac{1}{a^3 v'}\,((\,a^3v')\,'-a^5(\,m_X^2v-\frac{3}{4}\lambda\,v^2))\,. \label{phi'}
\end{eqnarray}
Here the prime denotes the derivative with respect to $z$.
We will firstly give a parametrization of the VEV $v(z)$ satisfying proper boundary conditions,
and then use the above equation to determine the background dilaton up to a integral constant.
We require that
\begin{eqnarray}
&&v\, \sim\,\, m_q\zeta z+\frac{\sigma}{\zeta}z^3\,,\quad  z\rightarrow0\,;\label{X0}\\[0.2cm]
&&v\, \sim\,\, \gamma z^2\,,\quad\hspace{1.4cm}  z\rightarrow\infty\,. \label{Xinf}
\end{eqnarray}
The constant $\zeta=\sqrt{3}/(2\pi)$. The condition at UV (i.e. $z\rightarrow0$) captures the effects of both
the explicit and spontaneous breaking of chiral symmetries. The condition at IR (i.e. $z\rightarrow\infty$)
is chosen for realizing an asymptotically linear nucleon spectra \cite{Z2}.
By using equation (\ref{phi'}) we have the following asymptotic behaviors of $\Phi(z)$
\begin{eqnarray}
&&\Phi\, \sim\, \frac{3}{4} \lambda\, m_q \zeta\, z + O(z^3)\,,\quad  z\rightarrow0\,;  \\[0.2cm]
&&\Phi\, \sim\, \frac{3}{16}\lambda \gamma\, z^2\,,\quad\hspace{1.5cm}  z\rightarrow\infty\,.
\end{eqnarray}
We can see that the effect of the dilaton can be neglected at the UV boundary, while in the deep
IR region it is quadratic which is needed for asymptotically linear spectra of mesons.
A simple parametrization of $v$ which satisfies both UV and IR boundary conditions, i.e.
(\ref{X0}) and (\ref{Xinf}), can be chosen as
\begin{eqnarray}
\frac{1}{2}\,v(z)=\frac{\,Az+Bz^3}{\sqrt{1+C^2z^2}\,}\,\,.  \label{VEV}
\end{eqnarray}
The relations between $m_q,\sigma,\gamma$ and $A,B,C$ are
\begin{eqnarray}
m_q=\frac{2A}{\zeta}\,,\qquad \sigma=2\zeta\,(B-\frac{1}{2}AC^2)\,,\qquad \gamma=\frac{2B}{C}\,.\label{rel}
\end{eqnarray}
The parameters $A,B,C$ and $\lambda$ are determined by fitting the experimental data of
the pseudoscalar, scalar, vector and axial-vector meson masses. We take their values as
\begin{eqnarray}
A=0.684\,\mathrm{MeV}\,,\quad
B=(300.74\,\mathrm{MeV})^3\,,\quad
C=1311.5\,\mathrm{MeV}\,,\quad
\lambda=27.24\,.\label{ABC}
\end{eqnarray}
Then by using (\ref{rel}) we can obtain the values of $m_q,\sigma,\gamma$ as
\begin{eqnarray}
m_q=4.96\,\mathrm{MeV}\,,\quad \sigma=(244.81\,\mathrm{MeV})^3\,,\quad \gamma=(203.66\,\mathrm{MeV})^2.
\end{eqnarray}
We will use them to calculate spectra and compare with the experimental data.

\subsection{Second order action}
To be more readable and fix our convention, in this subsection we review how to choose the gauge condition
and expand the action (\ref{SM}) to the quadratic order. Firstly we define
\begin{eqnarray}
X\,=\,\left(\frac{v}{2}+S\right)\,e^{2iP}\,,
\end{eqnarray}
where $S$ is a real scalar and $P$ is a real pseudoscalar. Also define
\begin{eqnarray}
V_M=\frac{1}{2}\,(L_M+R_M)\,,\quad
A_M=\frac{1}{2}\,(L_M-R_M)\,.
\end{eqnarray}
Expand the action (\ref{SM}) to the quadratic order of these new fields, we have
\begin{eqnarray}
S_M^{(2)}=\int d^4x\,dz \left(\,\mathcal{L}^{(2)}_{P,A_5}+\mathcal{L}^{(2)}_S+\mathcal{L}^{(2)}_{V}+\mathcal{L}^{(2)}_{A}\right)
\end{eqnarray}
with
\begin{eqnarray}
\hspace{-1.5cm}&&\mathcal{L}^{(2)}_{P,A_5} = -\frac{1}{2}\,a^3v^2e^{-\Phi}P^a\,\p^2P^a-
               \frac{1}{2}\,a^3v^2e^{-\Phi}(\p_5P^a-A_5^a)^2-\frac{1}{2g_5^2}\,ae^{-\Phi}A_5^a\,\p^{2}A_5^a\,,\\[0.2cm]
\hspace{-1.5cm}&&\mathcal{L}^{(2)}_S = -\frac{1}{2}\,a^3e^{-\Phi} S^a\left\{\p^{2}-
               \frac{1}{\,a^3e^{-\Phi}}\,\p_{5}(a^3e^{-\Phi}\p_5)+m_X^2a^2-\frac{3}{4}\lambda\,a^2v\right\}S^a\,,\label{LS}\\[0.2cm]
\hspace{-1.5cm}&&\mathcal{L}^{(2)}_{V} = -\frac{1}{2g_5^2}\,ae^{-\Phi}V_\mu^a\left\{-\eta^{\mu\nu}\p^2+\p^\mu\p^\nu+
               \frac{1}{ae^{-\Phi}}\,\p_5(ae^{-\Phi}\p_5)\eta^{\mu\nu}\right\}V_\nu^a\,,\label{LV}\\[0.2cm]
\hspace{-1.5cm}&&\mathcal{L}^{(2)}_{A} = -\frac{1}{2g_5^2}\,ae^{-\Phi}A_\mu^a\left\{-\eta^{\mu\nu}\p^2+\p^\mu\p^\nu+
               \frac{1}{ae^{-\Phi}}\,\p_5(ae^{-\Phi}\p_5)\eta^{\mu\nu}-g_5^2a^2v^2\eta^{\mu\nu}\right\}A_\nu^a\,.\label{LA}
\end{eqnarray}
Some cross-terms have been canceled by the following gauge fixing Lagrangian
\begin{eqnarray}
\mathcal{L}_{\mathrm{G.F.}}&=&-\frac{ae^{-\Phi}}{\,2g_5^2\,\xi_V}\left\{\p^{\mu}V_\mu^a-\frac{\xi_V}{\,ae^{-\Phi}}\,\p_5(ae^{-\Phi}V_5^a)\right\}^2\nonumber\\
&&\hspace{0.5cm}-\frac{ae^{-\Phi}}{\,2g_5^2\,\xi_A}\left\{\p^{\mu}A_\mu^a-\frac{\xi_A}{\,ae^{-\Phi}}\,\p_5(ae^{-\Phi}A_5^a)+g_5^2\,\xi_Aa^2v^2P^a\right\}^2\,.
\end{eqnarray}
We use the unitary gauge $\xi_{V,A}\rightarrow\infty$. This means that
\begin{eqnarray}
\p_5(ae^{-\Phi}V_5^a)&=&0\,,\\[0.2cm]
\p_5(ae^{-\Phi}A_5^a)&=&g_5^2\,a^3v^2e^{-\Phi}P^a\,.\label{AP}
\end{eqnarray}
The second equation allows us to write $P^a$ in terms of $A_5^a$. Therefore we can simplify $\mathcal{L}^{(2)}_{P,A_5}$ as follows
\begin{eqnarray}
\mathcal{L}^{(2)}_{P,A_5}\rightarrow\,\,\mathcal{L}^{(2)}_{A_5}=-\frac{1}{2g_5^2}\,ae^{-\Phi}A_5^a\,\p^2D^2A_5^a
     -\frac{1}{2}\,a^3v^2e^{-\Phi}(D^2A_5^a)(D^2A_5^a)\,,\label{LA5}
\end{eqnarray}
where the second order differential operator $D^2$ is defined by
\begin{eqnarray}
D^2f:=-\,\p_5\left(\frac{\p_5(ae^{-\Phi}f)}{\,g_5^2\,a^3v^2e^{-\Phi}}\right)+f\,.
\end{eqnarray}

\subsection{Pseudoscalar mesons}

To get the 4D effective action, we expand the field $A_5$ in terms of its Kaluza-Klein (KK) modes\footnote{From now
we drop the $su(2)$ index $a$ of all fields.}
\begin{eqnarray}
A_5(x,z)=\sum_{n=0}^{\infty}\,\pi^{(n)}(x)\,f_P^{(n)}(z)\,,\label{KKp}
\end{eqnarray}
with $f_P^{(n)}$ being the eigenfunction of the differential operator $D^2$
\begin{eqnarray}
-\,\p_5\left(\frac{\p_5(ae^{-\Phi}f_P^{(n)})}{\,g_5^2\,a^3v^2e^{-\Phi}}\right)+f_P^{(n)}=
\frac{M_P^{(n)2}}{\,g_5^2a^2v^2}\,f_P^{(n)}\,,\label{Mp}
\end{eqnarray}
with boundary condition \cite{DP2}
\begin{eqnarray}
\p_5(ae^{-\Phi}f_P^{(n)})|_{z\rightarrow0}=0\,,\quad\quad
f_P^{(n)}|_{z\rightarrow\infty}=0\,.\label{bcp}
\end{eqnarray}
According to general theories of the Sturm-Liouville problem,  we can normalize $f_P^{(n)}$
by the following orthonormality relation\footnote{If $m_q>0$, there is no zero mode solution, and the RHS makes sense.}
\begin{eqnarray}
\int_0^\infty \frac{e^{-\Phi}}{\,av^2}\,\,f_P^{(n)}f_P^{(n')}\,dz=\frac{g_5^4}{M_P^{(n)2}}\,\delta_{nn'}\,.
\end{eqnarray}
When insert (\ref{KKp}) into (\ref{LA5}) and do the integration over the $z$-coordinate, we get exactly
an effective 4D action for a tower of pseudoscalar fields $\pi^{(n)}$, which can be identified as the
pion and its radial excitations.
We can rewrite this eigenvalue problem in a Schr\"{o}dinger form. Define
\begin{eqnarray}
p=\frac{1}{\,g_5^2\,a^3v^2e^{-\Phi}}\,,\quad\quad
q=\frac{1}{\,ae^{-\Phi}}\,,
\end{eqnarray}
and $\psi_P^{(n)}=ae^{-\Phi}{p}^{1/2}f_P^{(n)}$, which satisfies the Schr\"{o}dinger equation
$-\psi_P^{(n)\prime\prime}+V_P\psi_P^{(n)}=M_P^{(n)2}\psi_P^{(n)}$ with the effective potential
\begin{eqnarray}
V_P=\frac{\,2pp''-p'^{2}+4pq}{4p^2}
\end{eqnarray}
It can be checked that $V_P\sim O(z^2)$ as $z\rightarrow\infty$ due to the background dilaton.
So the asymptotical spectrum is linear with respect to the radial quantum number $n$.
The eigenvalue problem (\ref{Mp}) and (\ref{bcp}) cannot be solved analytically.
We have to rely on numerical calculations. By using the parameters listed in (\ref{ABC}),
the resulting spectra and the corresponding experimental data are shown in Table \ref{pion}.

\begin{table}
\centering
\begin{tabular}{|c|c|c|c|c|c|}
\hline
$\pi$                   &  0    &   1  &  2   &  3   &  4  \\
\hline
$m_{\mathrm{th}}$ (MeV)  & 149 & 1239 & 1709 & 2072 & 2379 \\
\hline
$m_{\mathrm{ex}}$ (MeV)  & 139 & 1300 & 1816 & - & - \\
\hline
error                    & 7.2\% & 4.7\% & 5.9\% & - & - \\
\hline
\end{tabular}
\caption{\small{The theoretical and experimental values for pseudoscalar meson masses.}}\label{pion}
\end{table}

\subsection{Scalar mesons}

We expand the field $S$ in terms of its KK modes
\begin{eqnarray}
S(x,z)=\sum_{n=0}^{\infty}\,\phi^{(n)}(x)\,f_S^{(n)}(z)\,,\label{KKf}
\end{eqnarray}
where $f_S^{(n)}$'s are eigenfunctions of the following problem
\begin{eqnarray}
&&-\frac{1}{\,a^3e^{-\Phi}}\,\p_{5}(a^3e^{-\Phi}\p_5f_S^{(n)})+(\,m_X^2a^2-\frac{3}{4}\lambda\,a^2v)\,f_S^{(n)}=M_S^{(n)2}f_S^{(n)}\,,\nonumber\\[0.2cm]
&&\hspace{1cm} f_S^{(n)}|_{z\rightarrow0}=0\,,\quad\quad f_S^{(n)}|_{z\rightarrow\infty}=0\,.\label{Mf}
\end{eqnarray}
The orthonormality condition is
\begin{eqnarray}
\int_0^\infty a^3e^{-\Phi}f_S^{(n)}f_S^{(n')}dz=\,\delta_{nn'}\,.
\end{eqnarray}
Insert (\ref{KKf}) into (\ref{LS}) and do the integration over the $z$-coordinate,
we get exactly an effective 4D action for a tower of massive scalar fields $\phi^{(n)}$.
We can also transform (\ref{Mf}) into a Schr\"{o}dinger form, by setting
$f_S^{(n)}=e^{\omega_{_S}/2}\psi_S^{(n)}$ with $\omega_S=\Phi-3\log{a}$.
The effective potential $V_S$ for scalar mesons is
\begin{eqnarray}
V_S=\frac{1}{4}\omega_S'^{\,2}-\frac{1}{2}\omega_S''+m_X^2a^2-\frac{3}{4}\lambda\,a^2v\,.
\end{eqnarray}
Note that, due to the first term, the potential is $O(z^2)$ as $z\rightarrow\infty$.
Therefore the spectrum of scalar mesons is asymptotically linear.
Since there are several scalar mesons with relatively small mass, we take the sign
in front of the coupling constant $\lambda$ to be negative in order to fit the data better.
The theoretical and the experimental values of scalar mesons are listed in Table \ref{f0}.
The agreement is good except the first state, which is a light scalar with a very large width.
Since the parametrization (\ref{VEV}) of the VEV $v(z)$ is only the simplest one satisfying
boundary conditions (\ref{X0}) and (\ref{Xinf}), it seems possible that this discrepancy can be
improved by using a more complicated expression of $v(z)$.

\begin{table}
\centering
\begin{tabular}{|c|c|c|c|c|c|c|c|c|}
\hline
$f_0$                   &  0    &   1  &  2   &  3   &  4  &  5  &  6  &  7  \\
\hline
$m_{\mathrm{th}}$ (MeV)  & 118 & 953 & 1335 & 1627 & 1873 & 2089 & 2285 & 2465 \\
\hline
$m_{\mathrm{ex}}$ (MeV)  & 550 & 980 & 1350 & 1505 & 1724 & 1992 & 2103 & 2314 \\
\hline
error                   &78.6\%& 2.7\%&1.1\%& 8.1\%& 8.7\%& 4.9\%& 8.7\%& 6.5\% \\
\hline
\end{tabular}
\caption{\small{The theoretical and experimental values for scalar meson masses.}}\label{f0}
\end{table}

\subsection{Vector mesons}
We expand the field $V_\mu$ in terms of its KK modes
\begin{eqnarray}
V_\mu(x,z)=\sum_{n=0}^{\infty}\,\rho_\mu^{(n)}(x)\,f_V^{(n)}(z)\,,\label{KKr}
\end{eqnarray}
with $f_V^{(n)}$ being eigenfunctions of the following problem
\begin{eqnarray}
&&-\frac{1}{\,ae^{-\Phi}}\,\p_5(ae^{-\Phi}\p_5f_V^{(n)})=M_V^{(n)2}f_V^{(n)}\,,\nonumber\\[0.2cm]
&&\hspace{0.3cm} f_V^{(n)}|_{z\rightarrow0}=0\,,\quad\quad f_V^{(n)}|_{z\rightarrow\infty}=0\,.\label{Mr}
\end{eqnarray}
We normalize $f_V^{(n)}$ by the following orthonormality condition
\begin{eqnarray}
\int_0^\infty ae^{-\Phi}f_V^{(n)}f_V^{(n')}dz=\,\delta_{nn'}\,.
\end{eqnarray}
Insert (\ref{KKr}) into (\ref{LV}) and do the integration over the $z$-coordinate,
we get exactly an effective 4D action for a tower of massive vector fields $\rho_\mu^{(n)}$,
which can be identified as the fields of $\rho$ mesons.
We can also transform (\ref{Mr}) into a Schr\"{o}dinger form, by setting
$f_V^{(n)}=e^{\omega/2}\psi_V^{(n)}$ with $\omega=\Phi-\log{a}$.
The effective potential $V_V$ for vector mesons is
\begin{eqnarray}
V_V=\frac{1}{4}\omega'^{\,2}-\frac{1}{2}\omega''\,.
\end{eqnarray}
It is also of order $O(z^2)$ in the deep IR region, and gives us asymptotically linear spectra
for vector mesons.
The theoretical and the experimental values of vector mesons are listed in Table \ref{rho}.

\begin{table}
\centering
\begin{tabular}{|c|c|c|c|c|c|c|c|c|}
\hline
$\rho$                   &  0  &  1  &  2   &  3   &  4  &  5  &  6    \\
\hline
$m_{\mathrm{th}}$ (MeV)  & 759 & 1202 & 1519 & 1779 & 2005 & 2207 & 2393  \\
\hline
$m_{\mathrm{ex}}$ (MeV)  &775.5& 1282 & 1465 & 1720 & 1909 & 2149 & 2265  \\
\hline
error                   &2.1\%& 6.2\% &3.7\% & 3.4\%& 5.0\%& 2.7\%& 5.6\% \\
\hline
\end{tabular}
\caption{\small{The theoretical and experimental values for vector meson masses.}}\label{rho}
\end{table}

\subsection{Axial-vector mesons}

We expand the field $A_\mu$ in terms of its KK modes
\begin{eqnarray}
A_\mu(x,z)=\sum_{n=0}^{\infty}\,a_\mu^{(n)}(x)\,f_A^{(n)}(z)\,,\label{KKa}
\end{eqnarray}
with $f_A^{(n)}$ being eigenfunctions of the following problem
\begin{eqnarray}
&&-\frac{1}{\,ae^{-\Phi}}\,\p_5(ae^{-\Phi}\p_5f_A^{(n)})+g_5^2\,a^2v^2f_A^{(n)}=M_A^{(n)2}f_A^{(n)}\,,\nonumber\\[0.2cm]
&&\hspace{0.8cm} f_A^{(n)}|_{z\rightarrow0}=0\,,\quad\quad f_A^{(n)}|_{z\rightarrow\infty}=0\,.\label{Ma}
\end{eqnarray}
The orthonormality condition for $f_A^{(n)}$ is the same as vector mesons
\begin{eqnarray}
\int_0^\infty ae^{-\Phi}f_A^{(n)}f_A^{(n')}dz=\,\delta_{nn'}\,.
\end{eqnarray}
Insert (\ref{KKa}) into (\ref{LA}) and do the integration over the $z$-coordinate,
we get exactly an effective 4D action for a tower of massive axial-vector fields $a_\mu^{(n)}$,
which can be identified as the $a_1$ meson together with its radial excitation states.
We can also rewrite (\ref{Ma}) in a Schr\"{o}dinger form, by setting
$f_A^{(n)}=e^{\omega/2}\psi_A^{(n)}$ with $\omega=\Phi-\log{a}$.
The effective potential $V_A$ for axial-vector mesons is
\begin{eqnarray}
V_A=\frac{1}{4}\omega'^{\,2}-\frac{1}{2}\omega''+g_5^2\,a^2v^2\,. \label{VA}
\end{eqnarray}
It also has the same behavior at $z\rightarrow\infty$, and asymptotically linear spectra follows.
The theoretical and the experimental values of axial-vector mesons
are listed in Table \ref{a1}. From this table we can see the
discrepancy of the slope of the mass spectra. This is due to the
term $g_5^2\,a^2v^2$ in (\ref{VA}) which, unlike other soft-wall
models, e.g. \cite{GKK}, is also asymptotically quadratic at
infinity. This is a direct consequence of the asymptotic behavior
$v(z)$ in the deep IR region, which is crucial for realizing the
linear spectra in the nucleon sector. This fact seems to be a main
limitation of the present model.

\begin{table}
\centering
\begin{tabular}{|c|c|c|c|c|c|}
\hline
$a_1$                   &  0  &  1  &  2   &  3   &  4  \\
\hline
$m_{\mathrm{th}}$ (MeV)  & 997 & 1541 & 1934 & 2258 & 2540  \\
\hline
$m_{\mathrm{ex}}$ (MeV)  &1230 & 1647 & 1930 & 2096 & 2270  \\
\hline
error                   &18.9\%& 6.5\% &0.2\% &7.7\%& 11.9\% \\
\hline
\end{tabular}
\caption{\small{The theoretical and experimental values for axial-vector meson masses.}}\label{a1}
\end{table}

\section{Nucleon sector}

The spin-1/2 nucleon can be realized by introducing two Dirac spinors $\Psi_{1,2}$ in the bulk.
Each of them is also a isospin doublet, whose index is suppress for notational simplicity .
The action is
\begin{eqnarray}
S_N&=&\int d^5x\, \sqrt{G}\,\left(\mathcal{L}_K+\mathcal{L}_I\right) \,, \nonumber\\[0.1cm]
\mathcal{L}_K&=&i\overline{\Psi}_1\Gamma^M\nabla_M\Psi_1+i\overline{\Psi}_2\Gamma^M\nabla_M\Psi_2
                -m_{\Psi}\overline{\Psi}_1\Psi_1+m_{\Psi}\overline{\Psi}_2\Psi_2 \,, \\[0.2cm]
\mathcal{L}\,_I&=&-g\,\overline{\Psi}_1X\Psi_2-g\,\overline{\Psi}_2X^\dag\Psi_1\,.\label{SN} \nonumber
\end{eqnarray}
Here $\Gamma^M=e^M_A\Gamma^A=z\delta^M_A\Gamma^A$, and $\{\Gamma^A,\Gamma^B\}=2\,\eta^{AB}$
with $A=(a,5)$. We choose the representation as $\Gamma^A=(\gamma^a, -i\gamma^5)$ with
$\gamma^5=\mathrm{diag}(I,-I)$.
The covariant derivatives for spinors are
\begin{eqnarray}
\nabla_M \Psi_1&=&\p_M\Psi_1+\frac{1}{2}\,\omega^{AB}_M\Sigma_{AB}\Psi_1-iL_M\Psi_1 \,,\\
\nabla_M \Psi_2&=&\p_M\Psi_2+\frac{1}{2}\,\omega^{AB}_M\Sigma_{AB}\Psi_2-iR_M\Psi_2 \,.
\end{eqnarray}
Here $\Sigma_{AB}=\frac{1}{4}[\Gamma_A,\Gamma_B]$, and the nonzero components of the spin
connection $\omega^{AB}_M$ is $\omega^{a5}_\mu=-\omega^{5a}_\mu=\frac{1}{z}\,\delta^a_\mu$.
The parameter $m_\Psi$ is determined in principle by the dimension of the corresponding baryon
operator whose classical value is $\Delta_B=9/2$. Since it will receives quantum corrections
which cannot be calculated easily away from weak coupling limit, we treat $m_\Psi$ as a free
parameter, whose value will be determined by fitting the experimental data. The Yukawa term
$\mathcal{L}\,_I$ couples the Dirac spinors with the bulk scalar. It will give us the discrete
spin-1/2 nucleon spectra with a parity-doublet pattern. Therefore the behavior of the VEV of $X$
at infinity is crucial for asymptotically linearity of the nucleon spectrum. This requires
$v$ should be $O(z^2)$ as in (\ref{Xinf}).

\subsection{Nucleon spectra}

To study the spectrum it is enough to keep only the second order terms of $\Psi_{1,2}$
in the bulk action (\ref{SN}), those are
\begin{eqnarray}
S_N^{(2)}&=&\int d^5x\, \sqrt{G}\,\,(\,\mathcal{L}_K^{(2)}+\mathcal{L}_I^{(2)}) \,, \nonumber\\[0.1cm]
\mathcal{L}_K^{(2)}&=&\frac{1}{a}\,\sum_{i=1,2}\, \overline{\Psi}_i\left(i\gamma^\mu\p_\mu+
  \gamma^5\p_5+ \frac{2a'}{a}\gamma^5-m_{\Psi}a\right)\Psi_i\,,\\[0.1cm]
\mathcal{L}_I^{(2)}&=&\,-\frac{1}{2}\,gv\left(\,\overline{\Psi}_1\Psi_2+\overline{\Psi}_2\Psi_1\right)\,.\label{SN2} \nonumber
\end{eqnarray}
Similar to the meson sector, we expand $\Psi_{1,2}$ in terms of their KK modes
\begin{eqnarray}
\Psi_1(x,z)=\begin{pmatrix} \sum_n N_{L}^{(n)}(x)\,f_{1L}^{(n)}(z)\,\, \\[0.2cm] \sum_n N_{R}^{(n)}(x)\,f_{1R}^{(n)}(z)\,\,\end{pmatrix}\,\,,\quad
\Psi_2(x,z)=\begin{pmatrix} \sum_n N_{L}^{(n)}(x)\,f_{2L}^{(n)}(z)\,\, \\[0.2cm] \sum_n N_{R}^{(n)}(x)\,f_{2R}^{(n)}(z)\,\,\end{pmatrix}\,\,.\label{PsiKK}
\end{eqnarray}
Note that Dirac spinors in 5D contain four components, the same number as in 4D. In the above two equations,
$N^{(n)}_{L,R}$ are two-component objects, which will be interpreted as the left-handed and right-handed parts
of a tower of 4D nucleon fields, i.e. $N^{(n)}(x)=(N^{(n)}_L,N^{(n)}_R)^{\mathrm{T}}$ when reducing to a 4D
effective action. The four internal functions $f^{(n)}$ satisfy the following equations \cite{HIY}
\begin{eqnarray}
\begin{pmatrix} \p_z-\frac{\Delta^+}{z} & -u(z) \\[0.2cm] -u(z) & \p_z-\frac{\Delta^-}{z} \end{pmatrix}
\begin{pmatrix} f_{1L}^{(n)} \\[0.2cm] f_{2L}^{(n)} \end{pmatrix}&=&
-\,m_N^{(n)}\begin{pmatrix} f_{1R}^{(n)} \\[0.2cm] f_{2R}^{(n)} \end{pmatrix}\,\,, \label{EOM1}\\[0.3cm]
\begin{pmatrix} \p_z-\frac{\Delta^-}{z} & u(z) \\[0.2cm] u(z) & \p_z-\frac{\Delta^+}{z} \end{pmatrix}
\begin{pmatrix} f_{1R}^{(n)} \\[0.2cm] f_{2R}^{(n)} \end{pmatrix}&=&
+\,m_N^{(n)}\begin{pmatrix} f_{1L}^{(n)} \\[0.2cm] f_{2L}^{(n)} \end{pmatrix}\,\,, \label{EOM2}
\end{eqnarray}
with $\Delta^{\pm}=2\pm m_{\Psi}$, and
\begin{eqnarray}
u(z)=\frac{1}{2}\,\,g\,v(z)\,a(z)=\frac{\,g(A+Bz^2)}{\sqrt{1+C^2z^2}\,}\,\,. \label{u}
\end{eqnarray}
To get the correct chiral coupling between $\Psi$'s
and baryon operators at the UV boundary, we need to impose boundary conditions \cite{HIY}
\begin{eqnarray}
f_{1L}^{(n)}(z\rightarrow0)=0\,,\quad
f_{2R}^{(n)}(z\rightarrow0)=0\,. \label{bcf0}
\end{eqnarray}
For the boundary condition at infinity, we choose \cite{Z2}
\begin{eqnarray}
f_{1R}^{(n)}(z\rightarrow\infty)=0\,,\quad
f_{2L}^{(n)}(z\rightarrow\infty)=0\,.\label{bcfoo}
\end{eqnarray}
The orthonormality condition for the solution of this eigenvalue problem is
\begin{eqnarray}
\int_0^\infty a^4\left(f_{1L}^{(n)}f_{1L}^{(n')}+f_{2L}^{(n)}f_{2L}^{(n')}\right)dz=
\int_0^\infty a^4\left(f_{1R}^{(n)}f_{1R}^{(n')}+f_{2R}^{(n)}f_{2R}^{(n')}\right)dz=\,\,\delta_{nn'}\,.\label{ONn}
\end{eqnarray}
They can be proved by transforming (\ref{EOM1}) and (\ref{EOM2}) into a two-component
vector-valued Sturm-Liouville problem for $(f_{1L}^{(n)},f_{2L}^{(n)})^\mathrm{T}$ or
$(f_{1R}^{(n)},f_{2R}^{(n)})^\mathrm{T}$. Actually the boundary conditions (\ref{bcf0}) and (\ref{bcfoo})
are equivalent to requiring the normalizability of $f$'s.
From (\ref{EOM1}) and (\ref{EOM2}) it can be seen that only two of $f$'s are linear independent
\begin{eqnarray}
f_{2L}^{(n)}=-\epsilon f_{1R}^{(n)}\,,\quad\quad
f_{2R}^{(n)}=\epsilon f_{1L}^{(n)}\,,\label{P}
\end{eqnarray}
where $\epsilon=\pm1$ is the 4D parity. It can be checked that, by inserting (\ref{PsiKK}) into the second order action (\ref{SN2})
and do the integration over the $z$-coordinate, we obtain exactly an effective action for a tower of
4D Dirac spinor fields $N^{(n)}$ which are naturally identified as the nucleon $(p,n)$ and
its radial excitations.
We can also rewrite the problem in a coupled Schr\"{o}dinger form \cite{Z1}.
Define
$(f_{1L}^{(n)},f_{2L}^{(n)})=(z^2\chi_{1L}^{(n)},\,z^2\chi_{2L}^{(n)})$ and
$\chi^{(n)}_L=(\chi_{1L}^{(n)},\chi_{2L}^{(n)})^{\mathrm{T}}$, then
\begin{eqnarray}
-\chi_L^{(n)}{''}+V(z)\,\chi^{(n)}_L=\,m_N^{(n)\,2}\,\chi^{(n)}_L\,,\label{Sch}
\end{eqnarray}
where the potential matrix $V(z)$ is
\begin{eqnarray}
V(z)=\begin{pmatrix} V_{11} & V_{12} \\[0.2cm] V_{21} & V_{22} \end{pmatrix}
    =\begin{pmatrix} m_{\Psi}(m_{\Psi}-1)\,a^2+u^2 & u' \\[0.2cm] u' & m_{\Psi}(m_{\Psi}+1)\,a^2+u^2 \end{pmatrix}\,.\label{ptmat}
\end{eqnarray}
We also have similar equations for the right-handed fields.
Because of our parametrization (\ref{u}), the diagonal elements $V_{11}$ and $V_{22}$
are both of $O(z^2)$ as $z\rightarrow\infty$, and then the asymptotically linear nucleon spectra
follows form this fact.

Since equations (\ref{EOM1}) and (\ref{EOM2}) are already of first order,
it is convenient to directly solve them numerically. The parameter $g$ and $m_\Psi$
are determined by fitting the experimental data of the spinor-1/2 nucleon masses.
Their best values are
\begin{eqnarray}
g=13.9\,,\quad\quad
m_\Psi=1.3\,.\label{gm}
\end{eqnarray}
The resulting nucleon spectra and the corresponding data are listed in Table \ref{nucl}.

\begin{table}
\centering
\begin{tabular}{|c|c|c|c|c|c|c|c|}
\hline
$N$                     &  0     &   1    &  2     &  3     &  4     &  5     &   6  \\
\hline
$m_{\mathrm{th}}$(MeV)  & 939   & 1289   & 1441   & 1671  & 1802   & 1983   & 2100 \\
\hline
$m_{\mathrm{ex}}$(MeV) & 939    & 1440   & 1535   & 1650   & 1710   & 2090   & 2100 \\
\hline
error                   &0.0\%   &10.5\%   &6.1\%   &1.3\%   &5.4\%   &5.1\%   &0.0\% \\
\hline
\end{tabular}
\caption{\small{The experimental and theoretical values of the spin-1/2 nucleon masses.}}\label{nucl}
\end{table}

\subsection{Pion-Nucleon coupling}

We will discuss the pion-nucleon coupling in our soft-wall model.
It has been discussed in \cite{MT, KKY} that one of problems of the hard-wall AdS/QCD
model is that one need different IR cutoffs for mesons and nucleons. This becomes more
serious when discussing their couplings. In our soft-wall model, however, this problem
naturally disappear just by definition: the range of the $z$-coordinate is from 0 to
$\infty$ universally.

The coupling between the pion and the nucleon is dictated by the chiral symmetry.
The leading order interaction term is
\begin{eqnarray}
-ig_{\pi NN}\bar{N}\gamma^5\pi N\,.\label{pnn}
\end{eqnarray}
In the model of AdS/QCD, this coupling will comes from the covariant derivative terms and the Yukawa terms
in action (\ref{SN}), i.e.
\begin{eqnarray}
\mathcal{L}_{\mathrm{int}}(\Psi,A_5,P)=\,-\frac{i}{a}\,\overline{\Psi}\,\gamma^5A_5\sigma_3\Psi+gv\overline{\Psi}\,P\sigma_2\Psi\,,\label{Lpnn}
\end{eqnarray}
where $\Psi=(\Psi_1,\Psi_2)^{\mathrm{T}}$ and the Pauli matrices $\sigma_{2,3}$ act on this two-dimensional space.
$A_5=A_5^at^a$ and $P=P^at^a$ are the bulk fields for the pseudoscalar mesons. They are related through (\ref{AP}), i.e.
$g_5^2\,a^3v^2e^{-\Phi}P=\p_5(ae^{-\Phi}A_5)$. After inserting the expression (\ref{PsiKK}) into (\ref{Lpnn})
and compare with (\ref{pnn}), we have
\begin{eqnarray}
g_{\pi NN}&=&\int_0^\infty dz\,\,a^4\left[\,\left(f_{1L}^{(0)}f_{1R}^{(0)}-f_{2L}^{(0)}f_{2R}^{(0)}\right)f_P^{(0)}\right.\nonumber\\[0.2cm]
&&\hspace{2.2cm}\left.
-\frac{g\,e^\Phi}{\,g_5^2a^2v}\left(f_{1L}^{(0)}f_{2R}^{(0)}-f_{2L}^{(0)}f_{1R}^{(0)}\right)\p_5\left(ae^{-\Phi}f_P^{(0)}\right)\,\right]\,.
\end{eqnarray}
In the process of deriving this formula, we have used the parity relation (\ref{P}) and the orthogonality between the left-handed
and the right-handed spinors. The pion wave function $f_P^{(0)}$ is normalized by
\begin{eqnarray}
\int_0^\infty \frac{e^{-\Phi}}{\,av^2}\,\,f_P^{(0)\,2}\,dz=\frac{g_5^4}{M_P^{(0)2}}\,.
\end{eqnarray}
The nucleon wave functions are normalized by
\begin{eqnarray}
\int_0^\infty a^4\left(f_{1L}^{(0)\,2}+f_{2L}^{(0)\,2}\right)dz=
\int_0^\infty a^4\left(f_{1R}^{(0)\,2}+f_{2R}^{(0)\,2}\right)dz=1\,.
\end{eqnarray}
By using the parameters in (\ref{ABC}) and (\ref{gm}), the calculated value of the effective pion-nucleon coupling constant is
\begin{eqnarray}
g_{\pi NN}\simeq11.4\,.
\end{eqnarray}
Its value can be measured accurately, $g_{\pi NN}^{\mathrm{exp}}\simeq13.1$,
from the low energy pion-nucleon scattering. The agreement between the experimental value and our theoretical calculation
is rather good.

\section{Summary}

In this paper we study the soft-wall model involving both mesons and
nucleons extensively. We use only four parameters in our model to
fit more than twenty meson masses. The agreement between the
theoretical calculation and the experimental data is rather good,
with almost all errors within 10\%. By inheriting these meson
parameters, we use two more parameters to fit the spin-1/2 nucleon
masses. We also calculate the pion-nucleon coupling. To the author's
knowledge this is the first calculation of this quantity in a
soft-wall model. It is important that we can realize nucleons
properly in a soft-wall AdS/QCD model. The problem that mesons and
nucleons need different IR cutoffs in the hard-wall model disappears
here just by definition. It seems very natural to study
meson-nucleon interaction in this model.

\end{document}